\documentclass[aps,prl,twocolumn,superscriptaddress]{revtex4-2}

\usepackage{times}
\usepackage{amsmath,amssymb,bm}
\usepackage{graphicx}
\usepackage{float}
\newcommand{\Sv}{\mathbf{S}}
\newcommand{\Dv}{\mathbf{D}}
\newcommand{\jeff}{$j_{\rm eff}\!=\!1/2$}
\newcommand{\SIO}{Sr$_2$IrO$_4$}
\newcommand{\SIOb}{Sr$_3$Ir$_2$O$_7$}

\begin{document}

\title{Strain-driven spin flop and collapse of the giant magnon gap\\
in the bilayer iridate Sr$_3$Ir$_2$O$_7$}

\author{Choong H. Kim}
\email{choonghkim@ajou.ac.kr}
\affiliation{Department of Physics and Astronomy, Ajou University, Suwon, Korea}

\date{\today}

\begin{abstract}
The bilayer iridate \SIOb{} is a $c$-axis collinear antiferromagnet, held
there by a giant interlayer pseudodipolar anisotropy, whereas single-layer
\SIO{} cants in the $ab$ plane.  We show from first principles that biaxial
compression of a few percent ($\varepsilon_c\approx-2.4\%$) flops the easy
axis of \SIOb{} into the plane.  A magnetic model Hamiltonian built from
Wannier functions with no fitted parameter---reproducing the giant magnon
gap of the bulk, so far known only from fits to experiment---identifies the
mechanism.
Compression collapses the interlayer exchange channel, whose straight
Ir--O--Ir path weakens as the bent in-plane path strengthens.  Hund's
exchange sets the scale of the anisotropy and, beyond $J/U\approx0.15$,
removes the collinear state altogether.  The flop is not a rigid
rotation---the ordered moments of the two states cross at
$\varepsilon_c$---and it
carries a stark fingerprint, in that the giant easy-axis magnon gap
collapses to a gapless, Goldstone-like spectrum.  Compressively strained films thus sit on
a metamagnetic phase boundary ending in a zero-temperature bicritical point,
establishing epitaxial strain as a charge-neutral, symmetry-tailored handle
on spin--orbit-entangled order.
\end{abstract}

\maketitle

The cooperation of strong spin--orbit coupling (SOC) and moderate electron
correlation in $5d$ transition-metal oxides produces the \jeff{} Mott
insulators, in which magnetic moments are spin--orbital-entangled objects and
exchange interactions inherit an unusual bond-directional
character~\cite{Kim2009b,Jackeli2009,Moon2008,Bertinshaw2019,CaoSchlottmann2018}.  In the Ruddlesden--Popper
iridates this physics is laid out with rare clarity.  Beyond the dominant
isotropic exchange, the anisotropy is governed by pseudodipolar (PD)
interactions generated jointly by Hund's exchange and by the staggered
rotation of the IrO$_6$ octahedra~\cite{Jackeli2009,Jin2009,Kim2012}.  The
two members of the family answer to this anisotropy in opposite ways.
Single-layer \SIO{} orders as a canted antiferromagnet with moments locked in
the $ab$ plane~\cite{Kim2009b,Kim2012j,Wang2011d}, whereas bilayer \SIOb{}
is a collinear antiferromagnet with moments along
$c$~\cite{Kim2012,Cao2002}, and its magnon spectrum is gapped by
$\sim$90~meV---an anisotropy gap exceeding the magnon bandwidth---because the
interlayer PD coupling $\Gamma_c$ within the bilayer anchors the easy
axis~\cite{Kim2012a,Carter2013,MorettiSala2015}.  The dimensionality-driven change of easy
axis between the two compounds was identified early as a spin-flop transition
controlled by the ratio of intra- to interlayer PD
couplings~\cite{Kim2012}.

Switching the easy axis of \SIOb{} \emph{within one compound} is more than a
crystallographic curiosity.  The magnon gap, and with it the stability of the
proposed pseudospin $d$-wave state and its gap structure, is tied to the
moment direction~\cite{Kim2012a,Kim2014,Yan2015,He2015,KimdWave2016}; a controlled
reorientation would switch these properties without adding carriers.
Moreover, a system held just beyond a spin-flop instability realizes a
metamagnetic transition tunable by a $c$-axis field---the configuration that
produces field-tuned quantum criticality in the isostructural ruthenate
Sr$_3$Ru$_2$O$_7$~\cite{Grigera2001}.  Destabilizing the bilayer's robust
collinear order is in fact an actively pursued goal---by carrier
doping~\cite{Hogan2015}, hydrostatic pressure~\cite{Ding2016,ZhangJ2019},
and ultrafast optical driving~\cite{Mazzone2021}---but none of these
offers a static, continuous, charge-neutral handle on the easy axis
itself.  Epitaxial strain is that
knob.  It acts directly on the octahedral rotations and bond
geometry that generate the PD couplings, its power to reorganize
\jeff{} magnetism is established in the single-layer
compound~\cite{Lupascu2014,Serrao2014,Miao2014,Paris2020,Parschke2022}, and
heteroepitaxy of the layered iridates is well
developed~\cite{Matsuno2015,Hao2023}.  Density-functional work has already
established that compressive strain can drive such a flop in \SIOb{}, and
traced its energetics to the strain response of the intra- and interlayer
exchanges~\cite{Kim2017a,Kim2017}.  Building on that picture, what remains
to be resolved is the anisotropic part of the problem.  For the single-layer
compounds the anisotropic couplings have been derived ab initio, by
quantum-chemistry calculations~\cite{Katukuri2012,Katukuri2014} and by
constrained noncollinear DFT~\cite{LiuNCL2015}; for the bilayer, whose easy
axis hangs on the interlayer bond, they have so far been quantified only by
fitting model Hamiltonians to the measured magnon
spectrum~\cite{Kim2012a,Carter2013}.  Which interaction anchors the easy
axis and how it is lost under strain, what sets its size, and how the
transition would announce itself in experiment have therefore remained
open---a nontrivial task since the competing anisotropy energies are of meV
scale and originate in the interplay of Hund's coupling, SOC, and lattice
geometry.

\begin{figure}[tb]
\includegraphics[width=\columnwidth]{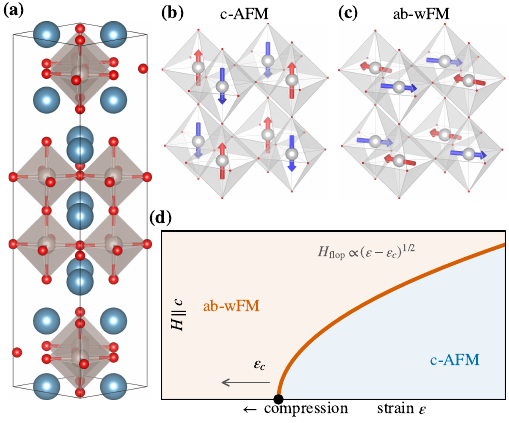}
\caption{Strain-driven spin flop in \SIOb{}.
(a) Crystal structure.  IrO$_6$ octahedra (O in red) share corners in plane
and the inner apical oxygen across a bilayer, with rock-salt SrO layers
between bilayers (Sr, large blue spheres); the octahedra are rotated by
$\pm\alpha$ about $c$ in a staggered pattern.
(b),(c) The two competing magnetic structures of the bilayer, arrows denoting
the \jeff{} moments.  In (b) the $c$-axis collinear antiferromagnet (c-AFM)
of the bulk, in (c) the $ab$-plane canted, weakly ferromagnetic state (ab-wFM)
reached beyond $\varepsilon_c$, whose Dzyaloshinskii--Moriya canting is
locked to the octahedral rotation.
(d) Schematic $T=0$ strain--field phase diagram.  Compression drives the flop
at $\varepsilon_c$, and on the collinear side a $c$-axis field completes it
along a metamagnetic line
$H_{\rm flop}\propto(\varepsilon-\varepsilon_c)^{1/2}$ ending at a
bicritical point (see text).}
\label{fig:idea}
\end{figure}

Here we show from first principles that biaxial compression of a few percent
turns the easy axis of \SIOb{} into the plane, and we identify the
interaction whose collapse does it.  Figure~\ref{fig:idea} lays out the idea.
The calculations for (001)-strained films give a spin flop from the $c$-axis
collinear state to the $ab$-plane canted, weakly ferromagnetic state, which
places compressively strained \SIOb{} on a metamagnetic phase boundary.  A
magnetic model derived from the calculated electronic structure with no
fitted parameter puts the anisotropic Hamiltonian of the bilayer iridate on
a first-principles footing---it reproduces the giant magnon gap at the
correct scale---and traces the flop to the collapse of the interlayer
exchange channel, whose straight Ir--O--Ir path weakens as the $c$ axis
expands and so erodes the pseudodipolar anisotropy that anchors the $c$
axis.  Hund's exchange, the
origin of that anisotropy~\cite{Jackeli2009,Kim2012}, sets its scale and
beyond $J/U\approx0.15$ removes the collinear state altogether.  The flop is
also not a rigid rotation.  The ordered moments of the two states become
equal at $\varepsilon_c$, so the moment changes length as it turns---and the
spectrum registers the turn as the collapse of the giant easy-axis magnon gap
into a gapless branch.

\begin{figure}[tb]
\includegraphics[width=\columnwidth]{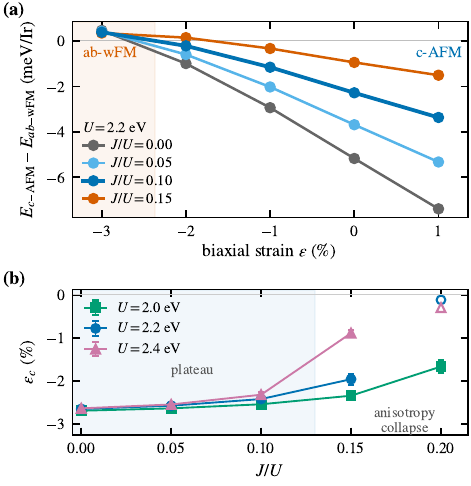}
\caption{Spin flop and its Hund's-coupling control.
(a) Total-energy difference
$\Delta E=E_{\rm c\mbox{-}AFM}-E_{ab\mbox{-}wFM}$ per Ir versus biaxial
strain at $U=2.2$~eV for four values of $J/U$.  The sign change marks the
flop, at $\varepsilon_c=-2.4\%$ for $J/U=0.10$, and shading marks the
$ab$-wFM side of that series.  (b) Critical strain versus $J/U$ for three values of $U$.  Bars give the
spread between the two ways of locating the sign change, and
open symbols mark series in which the collinear state is not the ground
state anywhere in the window.}
\label{fig:deltaE}
\end{figure}

Calculations use OpenMX~\cite{openmx} with a PBEsol~\cite{pbesol}
exchange--correlation functional, implemented here as a local patch;
norm-conserving pseudopotentials and pseudo-atomic orbitals,
noncollinear spinors with SOC, and the rotationally invariant DFT$+U$
(Lichtenstein) functional~\cite{RyeeHan2018} with spin-dependent
fully-localized-limit double counting; $U=2.2$~eV and $J=0.22$~eV on Ir
$5d$ unless scanned.  The cell is the 24-atom primitive $Bbcb$ bilayer cell
with $8\times8\times4$ $k$ points and a 300~Ry mesh cutoff.  Structures are
fixed by nonmagnetic scalar-relativistic relaxations at each strain, with
the in-plane lattice clamped to $a=a_0(1+\varepsilon)$
($a_0=5.522/\sqrt{2}$~\AA{} pseudocubic) while $c/a$ and the internal
coordinates are optimized.  Both magnetic states are then computed as SCF
solutions on these frozen structures, initialized along $c$ and in plane,
which keeps the meV-scale comparison between them free of structural bias.
An independent \textsc{vasp} calculation confirms the strain-driven spin flop with
qualitatively consistent results.
Magnon spectra are obtained by linear
spin-wave theory on the model couplings, with a Bogoliubov (Colpa) diagonalization of the four-sublattice bilayer problem.
The superexchange construction, the classical and spin-wave analyses, and the
tests on which the robustness of the transition rests are described in the
Supplemental Material (SM)~\cite{SM}.

Figure~\ref{fig:deltaE}(a) shows the total-energy difference
$\Delta E=E_{\rm c\mbox{-}AFM}-E_{ab\mbox{-}wFM}$ per Ir.  At zero strain the
collinear state is favored by $2.3$~meV/Ir, consistent with the bulk order;
under biaxial compression $\Delta E$ rises almost linearly
($\simeq0.9$~meV/Ir per \%) and changes sign at
$\varepsilon_c\simeq-2.4\%$ for $(U,J)=(2.2,0.22)$~eV.  The transition
announces itself on both sides, the collinear state tilting progressively
away from $c$ as $\varepsilon_c$ is approached and the in-plane state
carrying a growing net moment ($\sim0.1\,\mu_B$/Ir) from the
Dzyaloshinskii--Moriya canting.

The role of the Hund's coupling is summarized in Fig.~\ref{fig:deltaE}(b),
and it is twofold.  At fixed $U$, increasing $J$ suppresses $|\Delta E|$ at
every strain---the anisotropy energy at $\varepsilon=0$ falls from $5.3$ to
$0.6$~meV/Ir between $J=0$ and $0.4$~eV---yet the critical strain barely
moves.  It occupies a plateau, drifting only from $-2.7$ to
$-2.3\%$ over $J/U=0$--$0.125$ for all three values of $U$.  Beyond
$J/U\approx0.15$ the plateau ends abruptly, $\varepsilon_c$ collapsing toward
zero and, by $J/U=0.2$, the collinear state is no longer the ground state
anywhere in the studied strain window [open symbols in
Fig.~\ref{fig:deltaE}(b)].  The existence of the flop is thus a robust
prediction, robust against the correlation parameters within their physical
range, while the same Hund's exchange that generates the pseudodipolar
anisotropy~\cite{Jackeli2009,Kim2012} sets its scale and the threshold at
which it fails.

To identify what the strain actually does, we construct at each strain a
$t_{2g}$ Wannier Hamiltonian from the nonmagnetic, SOC-free calculation and
convert its bond matrices and measured tetragonal splitting into exchange
couplings by exact diagonalization of a two-site Kanamori$+$SOC cluster,
projecting onto the \jeff{} doublets and decomposing the resulting
$4\times4$ Hamiltonian exactly on the complete operator basis~\cite{SM}.  The
procedure contains \emph{no fitted parameter}, since hoppings, crystal
fields, and rotations are those of the actual strained structures.  For each bond it
yields the couplings of the established minimal
model~\cite{Jackeli2009,Kim2012,Carter2013},
$\mathcal{H}_{ij}=J_{ij}\,\Sv_i\!\cdot\!\Sv_j+\Gamma_{ij}S^z_iS^z_j
+\Dv_{ij}\!\cdot\!(\Sv_i\times\Sv_j)$.

\begin{figure}[tb]
\includegraphics[width=\columnwidth]{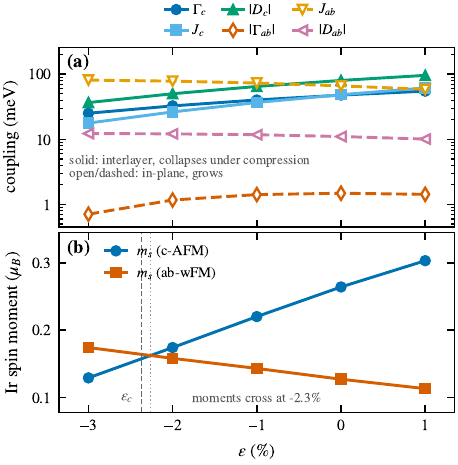}
\caption{Mechanism of the flop.  (a) All six nearest-neighbour couplings
versus strain from the Wannier-based two-site model ($\eta=J_H/U=0.10$;
logarithmic scale); filled symbols and solid lines are the interlayer bond,
open symbols and dashed lines the in-plane bond.  (b) Ordered Ir spin
moments of the two states from first principles, crossing at
$\approx\varepsilon_c$ (dashed line).}
\label{fig:mech}
\end{figure}

Figure~\ref{fig:mech}(a) shows the result.  Under compression the entire
interlayer channel collapses---$J_c$, $\Gamma_c$, and $|D_c|$ all fall by a
factor $2$--$3$ across the strain window---while the in-plane pseudodipolar
coupling $\Gamma_{ab}$ stays two orders of magnitude below $\Gamma_c$
throughout and the in-plane $J_{ab}$ grows.  The
origin is geometric.  The interlayer Ir--O--Ir path through the inner apical
oxygen is straight ($180^\circ$) and weakens as the $c$ axis expands under
biaxial compression, whereas the in-plane path, bent by the octahedral
rotation, strengthens as the in-plane bonds shorten.  Since the collinear
state is anchored almost entirely by $\Gamma_c$~\cite{Kim2012a,Carter2013},
its anisotropy advantage $A=S^2(2\Gamma_{ab}+\Gamma_c/2)$ erodes
continuously---no level crossing or sign change of any individual coupling
is required.  Resolving this channel separately is what locates the
transition.  The isotropic exchanges reorganize under strain as
well~\cite{Kim2017a}---in our couplings $J_{ab}$ overtakes $J_c$ near
$\varepsilon=+0.9\%$---but the moment direction is fixed by $\Gamma_c$ alone,
and its erosion places the reorientation more than $3\%$ away in strain.
Within the strain window of the transition the model tracks
the first-principles energetics quantitatively---the strain slope of its
anisotropy balance matches the DFT slope to within a few percent, and it is
as insensitive to $J$ as the DFT plateau of
Fig.~\ref{fig:deltaE}(b)~\cite{SM}.  Deeper compression lies outside
the model's language, because beyond $\varepsilon\approx-4\%$ the tetragonal
splitting becomes comparable to the SOC scale and the ground doublet departs
from \jeff{}, so all statements here are confined to the window around the
transition~\cite{SM}.

\begin{figure}[tb]
\includegraphics[width=\columnwidth]{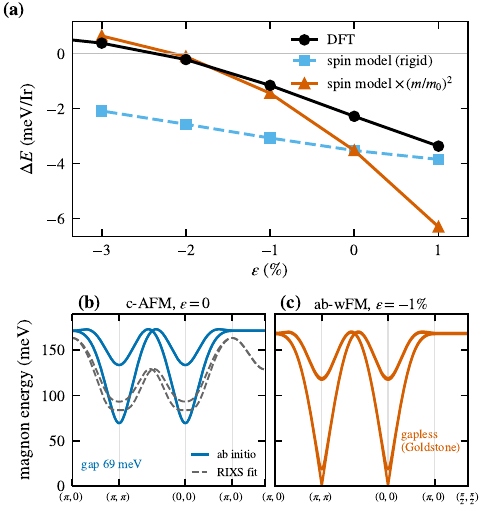}
\caption{Model versus first principles, and the spectroscopic fingerprint.
(a) $\Delta E(\varepsilon)$ from first principles (circles), from the
rigid-pseudospin model (squares), and from the model with its anisotropy
scaled by the calculated moments, $(m/m_0)^2$ (triangles).
(b) Linear spin-wave spectrum of the collinear state at $\varepsilon=0$ from
the Wannier-derived couplings (solid), compared with the closed-form
dispersion of the couplings fitted to RIXS~\cite{Kim2012a} (dashed).
(c) Spectrum of the in-plane state at $\varepsilon=-1\%$, anisotropy
renormalized by the calculated moments (SM); the acoustic branch is
gapless.}
\label{fig:model}
\end{figure}

A rigid-pseudospin picture, however, is not the whole story.
Figure~\ref{fig:mech}(b) shows the calculated ordered moments of the two
states.  Under compression the collinear moment collapses (by $55\%$ across
the window) while the in-plane moment grows (by $53\%$), and the two cross
at $\varepsilon\simeq-2.3\%$---essentially at $\varepsilon_c$.  The
insulating gap of the collinear state narrows in parallel, and part of the
moment evolution is a wavefunction effect.  As the tetragonal field changes
sign under compression, the $g$ tensor of the \jeff{} doublet inverts,
reducing the moment the $c$-oriented state can display while enhancing the
in-plane one; the magnitude of the calculated asymmetry, however, exceeds
this single-ion effect and points to O-$2p$ covalency beyond the $t_{2g}$
description~\cite{SM}.
A soft, composition-dependent moment rather than a rigid pseudospin also
resonates with the quantum-dimer and excitonic-insulator descriptions of
\SIOb{} from RIXS~\cite{MorettiSala2015,Mazzone2022}.  This
state-dependent moment softening is invisible to any model that assigns both
states the same rigid $S=1/2$, and it matters quantitatively.  The rigid
model, for all its success with the anisotropy balance, underestimates the
slope of $\Delta E(\varepsilon)$, while scaling its anisotropy terms by the
calculated moments, $(m/m_0)^2$, restores the missing strain dependence and
brackets the first-principles $\varepsilon_c$
[Fig.~\ref{fig:model}(a)].  Direction and magnitude of the moment therefore
evolve together, and the transition falls outside the standard spin-flop
picture of a rigid moment rotating in a fixed anisotropy landscape.

The same couplings determine the spin dynamics, computed here by linear
spin-wave theory [Fig.~\ref{fig:model}(b,c)].  For the unstrained collinear
state the calculation confronts experiment directly, yielding with no fitted
parameter an anisotropy gap of $69$~meV at $(\pi,\pi)$ and a band
top of $173$~meV [Fig.~\ref{fig:model}(b)], to be compared with the
resonant-inelastic-x-ray-scattering (RIXS) dispersion, whose fitted model
gives a $\approx92$~meV gap and a $\approx160$~meV band
top~\cite{Kim2012a}.  The gap-exceeds-bandwidth structure
characteristic of bilayer iridate magnetism thus emerges at the correct
scale, and the residual differences in dispersion shape trace to the
longer-range exchanges retained in the experimental
fit~\cite{Kim2012a} but not here.  On the
compressed side of the transition, the same machinery---with the
anisotropy renormalized by the calculated moments~\cite{SM}---gives a
dynamically stable in-plane spectrum
whose acoustic branch is gapless [Fig.~\ref{fig:model}(c)].  The model
anisotropy is uniaxial about $c$, so the in-plane state retains a U(1)
degeneracy and its Goldstone mode.  (The rigid model has no stable in-plane
spectrum~\cite{SM}.)  In the material this
degeneracy is broken only by weak higher-order in-plane terms, so
``gapless'' becomes a gap of at most a few meV---against $69$~meV on the
collinear side, an order-of-magnitude collapse.  The spin flop should
therefore announce itself spectroscopically, and RIXS or Raman scattering on
compressively strained films would see the defining gap of \SIOb{}
collapse across $\varepsilon_c$.

Figure~\ref{fig:idea}(d) summarizes the resulting zero-temperature phase
diagram.  For compression short of $\varepsilon_c$ the collinear state
survives with a strain-thinned anisotropy barrier, and a $c$-axis field
completes the flop at
$g\mu_BH_{\rm flop}\simeq(4\bar{z}\bar{J}\,|\Delta E|)^{1/2}
\propto(\varepsilon-\varepsilon_c)^{1/2}$, where
$\bar{z}\bar{J}\approx0.3$~eV is the summed isotropic exchange.  Because
that exchange is large the metamagnetic line is steep, and the flop field falls
from an inaccessible $\sim\!10^2$-T scale in the bulk to laboratory fields
only within a few hundredths of a percent of $\varepsilon_c$, so the field
is an ultrafine knob on top of the coarse strain axis, and the practical
routes to the boundary are substrate choice and continuous strain
tuning~\cite{Hicks2025}.
The terminus $(\varepsilon_c,H=0)$ is a zero-temperature bicritical point
at which easy-axis and easy-plane orders become degenerate and the magnon
gap collapses [Fig.~\ref{fig:model}(b,c)]---an insulating,
spin-orbit-entangled counterpart of the metamagnetic criticality of the
isostructural metallic ruthenate Sr$_3$Ru$_2$O$_7$~\cite{Grigera2001}.
The soft anisotropy landscape there is also a natural home for
fluctuation-stabilized phases of the kind recently found in the
square-lattice iridates~\cite{KimNematic2024}.  The reorientation also switches the giant easy-axis magnon
gap~\cite{Kim2012a} and, through it, the proposed pseudospin $d$-wave gap
structure tied to the moment direction~\cite{Kim2014,Yan2015}.  On the
materials side the required compression is demanding but available.
Taking the pseudocubic lattice constant $a_0\simeq3.90$~\AA, LSAT and
NdGaO$_3$ ($-0.7$ to $-1.1\%$) remain safely on the collinear side while
LaAlO$_3$ ($-2.8\%$) falls inside the predicted range, so the standard
substrates already bracket the flop---the transition and its mechanism, not
a critical strain known to a tenth of a percent.  More broadly, our results promote the Hund's
coupling from a parameter of model Hamiltonians to a design variable of
iridate magnetism, since it sets not only the size of the anisotropy but where a
film on a given substrate sits relative to the spin-flop instability.
Strain thereby joins the emerging program of using symmetry-tailored elastic
fields to reveal and control spin--orbit-entangled
orders~\cite{Pourovskii2025}.

\begin{acknowledgments}
This work was supported by National Research Foundation of Korea (NRF)
grants funded by the Korean government: RS-2026-25491029 (MSIT),
RS-2021-NR060141 (Ministry of Education), and the Global--LAMP Program
grant RS-2023-00285390 (Ministry of Education).
\end{acknowledgments}

\bibliography{Sr3Ir2O7-spinflop-letter}

\end{document}